\documentclass[twocolumn,preprintnumbers,amsmath,amssymb,pre]{revtex4}

\usepackage{graphicx}

\begin{document}

\preprint{}

\title{Self-organizing social hierarchies on scale-free networks}

\author{Lazaros K. Gallos}

\affiliation{Department of Physics, University of Thessaloniki, 54124 Thessaloniki, Greece
\\
gallos@physics.auth.gr}

\date{\today}

\begin{abstract}
In this work we extend the model of Bonabeau {\it et al.} in the case of scale-free networks.
A sharp transition is observed from an egalitarian to an hierarchical society,
with a very low population density threshold. The exact threshold value also depends
on the network size. We find that in an hierarchical society the number of individuals
with strong winning attitude is much lower than the number of the community members that
have a low winning probability.
\end{abstract}


\maketitle

\section{Introduction}

Self-organization of society structures and the formation of hierarchies
has always been an important issue in sociological studies.\cite{landau,chase} Recently, a
fresh point of view in the same problem was introduced through application
of statistical physics concepts and methods. A simple, yet powerful enough,
model was introduced by Bonabeau {\it et al.}\cite{bonabeau} in order to explain the uneven
distribution of fame, wealth, etc. The model was further modified later by
Stauffer,\cite{Stauffer1,Stauffer2} who introduced a feedback mechanism for determining the probability
of one's social rise or fall in the hierarchy.

The above model places the interacting individuals on a lattice, so that the
space, as experienced by a participant, is homogenous. Recently, though, a huge number
of observations on social (among many others) systems has revealed a strongly
inhomogeneous character in the number of connections between individuals.\cite{Albert,Dorogovtsev} In the
present study, thus, we extend the model of Bonabeau {\it et al.} in the case where the
substrate of the agents motion and interaction is such a scale-free network.

\section{The Model}

In the original version of the model proposed by Bonabeau {\it et al.}\cite{bonabeau}
a number of agents are distributed randomly on a $L\times L$ lattice, occupying a concetration $p$
of the total number of lattice sites.
Each site can host only one individual. These individuals perform isotropic random walks on
the lattice. A random agent is chosen and moves equiprobably to one of its four neighboring
sites, while the system time advances by $1/pN$ (when all individuals have moved on average once,
time is considered to have advanced by one Monte-Carlo step). Each person $i$ is characterized by a parameter $h_i$ which is a measure of
an individual's `fitness' and can represent wealth, power, or any property that is judged to be
important in a society. Initially all participating agents are of equal status ($h_i=1$) and
there is no hierarchy in the population. When in the process of the random walk, though,
an individual $i$ tries to visit a site already occupied by another person $j$,
there is a fight between the two. If the `attacking' person wins then $i$ and $j$ exchange
their positions. Otherwise, they remain in their original sites.
The outcome of the fight depends on the `strength' $h$ of the two opponents, with a probability
$q$ that $i$ wins over $j$:
\begin{equation}
q = \frac{1}{1+\exp\left[ \eta (h_j - h_i ) \right]} \,,
\end{equation}
where $\eta$ is a free parameter, with a constant value within each realization. After a
fight the fitness $h$ of a person participating in a fight is updated: the fitness of
the winner $h$ increases by 1, while the fitness of the loser decreases by 1. Thus, the
variable $h_i$ measures the number of wins minus the number of losses, but it is also
modified by an effect of fading memory. After one Monte Carlo step the fitness of all individuals
decreases to 90\% of its current value. In other words, in order to keep a large enough strength
it is not enough to have won a lot of fights in the past and remain inactive, but one must always retain one's strength
by participating (and winning) in fights. When the density of participants is low, this memory loss
is the prevailing mechanism that drives the system towards the egalitarian status, since fights
in that case are rare.

The level of separation in a society is measured via an order parameter, which is taken to be
the dispersion in the probability of winning a fight
\begin{equation}
\sigma = \left( \left\langle q^2 \right\rangle - \left\langle q \right\rangle^2 \right)^{1/2}  \,.
\end{equation}
The average is considered over all fights occuring within one Monte Carlo time step. A large
value of $\sigma$ reveals an hierarchical society where the probability of winning differs
considerably among the population. On the contrary, values of $\sigma$ close to zero imply
that on the average all society members `fight' each other in terms of equivalent forces.

In the original paper a phase transition was observed upon increasing the density $p$,
from $\sigma=0$ to a finite $\sigma$ value. Sousa and Stauffer,\cite{Stauffer1} though, pointed out that
the transition was an artifact of the simulations and this transition was in fact absent.
Later, Stauffer proposed a different mechanism for calculating the winning probability,\cite{Stauffer2}
where feedback from the current system state was introduced in the following form:
\begin{equation}
q = \frac{1}{1+\exp\left[ \sigma (h_j - h_i ) \right]} \,.
\label{EQ_Stauffer}
\end{equation}
In this case, an hierarchically organized population (large $\sigma$ value) enhances the probability
of the strongest member to win, and thus introduces a preference towards already strong individuals.
This mechanism restored the sharp transition of $\sigma$ with increasing $q$, yielding a critical
value close to $q_c=0.32$.

In this work we apply the modified model of Eq.~(\ref{EQ_Stauffer}) on scale-free networks.
A scale-free network is a graph where the probability that a node has $k$ links
to other nodes follows a power law
\begin{equation}
P(k) \sim k^{-\gamma} \,,
\label{Pk}
\end{equation}
where usually $2<\gamma<4$. We prepare a number of different networks (typically
of the order of 100) with a given $\gamma$ value using the configuration model:
First, the number of links for each node is determined by drawing random numbers
from the distribution (\ref{Pk}) and then links are established between randomly
chosen pairs of nodes. Care is taken to avoid self-links or double links between
two nodes. This process may create isolated clusters of nodes, so in our
simulations we only keep the largest cluster in the system which (depending on
$\gamma$) comprises 35-100\% of the number of system nodes $N$.

Individuals are randomly placed on the system nodes and move along the links. A person
on a node with $k$ connections choses randomly one of the connected nodes with probability
$1/k$ and tries to jump there. If the node is occupied a fight takes place under the
same rules as in the case of the lattice.

\section{Results}

\begin{figure}
\begin{center}
\includegraphics{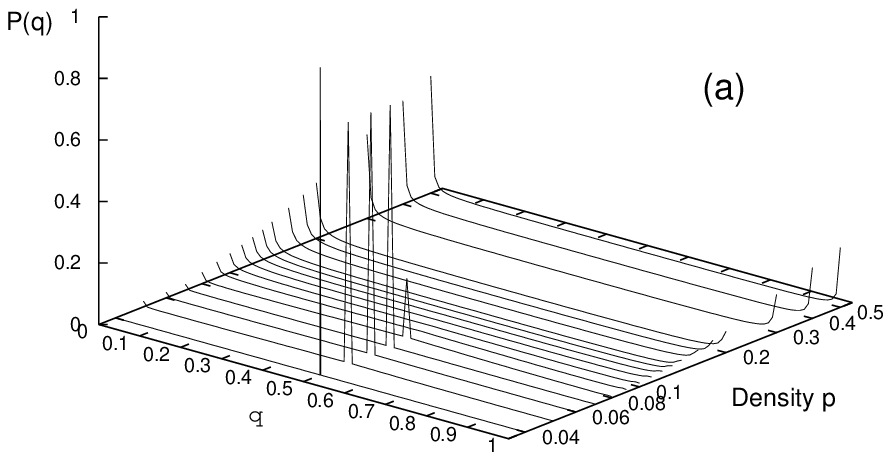}
\includegraphics{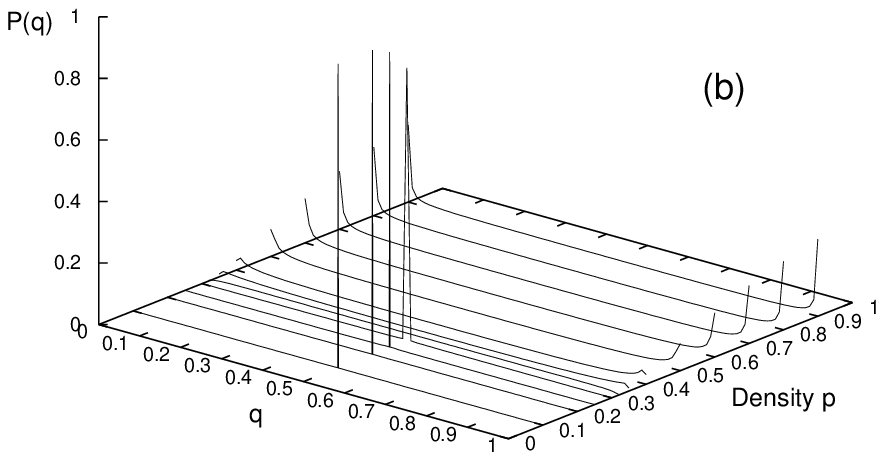}
\end{center}
\caption{Evolution of the distribution $P(q)$ with increasing density $p$ of the population,
(a) on the largest cluster of a network with $\gamma=3.0$ and $N=10^5$ nodes,
and (b) on regular two-dimensional lattice. \label{fig1}}
\end{figure}

In Fig.~\ref{fig1}(a) we present the distribution of $q$ for different population densities
$p$, for networks with $\gamma=3.0$, and $N=10^5$ nodes. We have verified that the observed
behavior is in general valid for other values of $\gamma$ as well.
When $p$ is very small, there are only rare encounters between the individuals and all winning
probabilities are equal
to $0.5$, which yields a delta function distribution up to $p=0.04$. When $p$ becomes greater
than $p=0.05$ the form of the distribution changes drastically. The peak is getting lower,
until it completely dissapears. Now, in the region around $p=0.1$ all winning probabilities are almost
equiprobable and evenly distributed among the population. Upon further increasing $p$ a strong
polarization arises in the population with most people having a vanishing winning probability.
Very few individuals have intermediate values of $q$, and another peak appears in the distribution
close to the area of complete dominance $q=1$. The intensity of this peak is lower than the peak at $q=0$,
indicating that the clique of `strong' individuals has fewer members than the community of
low-`strength' members.

\begin{figure}
\begin{center}
\includegraphics{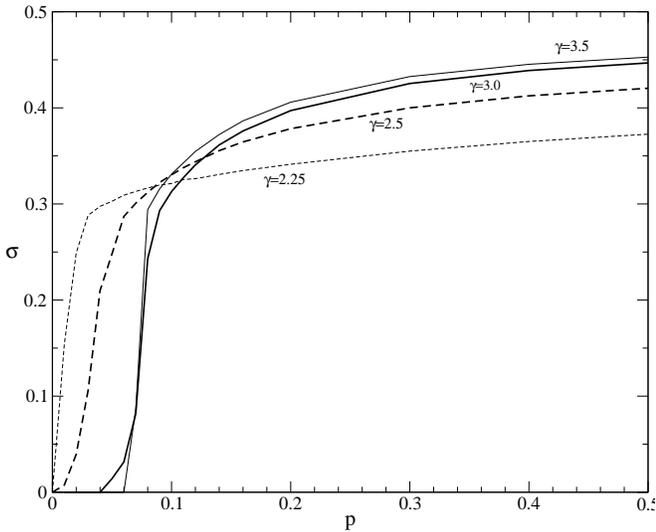}
\end{center}
\caption{Order parameter $\sigma$ as a function of the population density $p$ for scale-free
networks with exponent $\gamma=2.25$, $2.5$, $3.0$, and $3.5$ (shown on the plot). Results were averaged over
100 different network realizations of $N=10^5$ nodes, using typically $10^5$ steps per run. \label{fig2}}
\end{figure}

Comparison with the case of a lattice (shown in Fig.~\ref{fig1}(b)) reveals some interesting features.
The general behavior
is similar (going from a delta function to uniform distribution to increasing peaks at the edges of
the distribution). However, the range over where these transitions take place is very different,
with networks leaving the egalitarian state in much lower densities (notice the logarithmic axis
of $p$ in Fig.~\ref{fig1}(a)). More important is the observation that on a lattice the two peaks
of the winning probability distribution at high population densities are completely symmetric.
This symmetry is due to the homogeneity of the lattice, contrary to the result for the scale-free
networks. On a network, when an individual with high winning probability is placed on a hub
will fight against many opponents, who have lower $q$. These low-$q$ individuals at the branches of
the hub try to pass through the hub, where they will probably lose the fight. In this
way, they will become weaker while they will further strengthen the already strong person.
In practice, one strong individual can keep a quite large number of weaker people into a losing
state, which is the mechanism underlying the observed assymetry in the two peaks. In general,
roughly 60-65\% of the individuals belong to the $q\sim 0$ community and 20-25\% belong to the $q\sim 1$ clique.

The observed change in the distribution shape with increasing $p$ already hints the existence
of a sharp phase transition. This transition is indeed verified by our simulations,
when using our order parameter $\sigma$. The results are presented in figure \ref{fig2}.

The critical threshold for all $\gamma$ values is significantly lower than in the case of
lattices (where $p_c=0.32$). In fact, when $\gamma=2.25$ or $\gamma=2.5$ there is almost no
threshold and an hierarchical society emerges as soon as there is a non-zero population on the lattice,
due to the frequent encounters. For $\gamma\geq 3$ the threshold has a finite value, which is still
in low densities, of the order of $0.05$. It is also noteworthy that for networks with low $\gamma$,
the asymptotic value of the order parameter is smaller than the one for networks with large $\gamma$ values.
This shows that the most-connected networks initially establish an hierarchical society at lower densities
than less connected networks, but retain a lower level of hierarchy at larger densities.

The network heterogeneity also introduces another effect, apart from moving
the critical threshold closer to $p=0$. For concentrations close to criticality from below,
the behavior of $\sigma$ in networks of the same $\gamma$ may be very different.
Thus, for e.g. $\gamma=3.0$ and $p=0.5$, in most realizations the value $\sigma$
vanishes in a few hundrend steps. In a significant percentage (roughly 10-15\%)
of the realizations, though, we have observed that $\sigma$ would retain a large value and
fluctuate around $\sigma=0.3$ even after $10^4$ steps.
Inspection of these realizations revealed that almost all of them finally converge to $\sigma=0$,
but the time for equilibration may be of the order of $10^6$ steps or even more, while the fluctuations
in the value of $\sigma$ with time are large ($\sigma$ can assume values very close to 0 or rise up to 0.45,
before settling to $\sigma=0$).
These long relaxation times and the wide dispersion of $\sigma$ show that a society with a density close to the
criticality on a scale-free network may remain in turbulence for a long time, and even a small number
of individuals may separate into different hierarchies for a significant duration, before finally settling to an
egalitarian society.

\begin{figure}
\begin{center}
\includegraphics{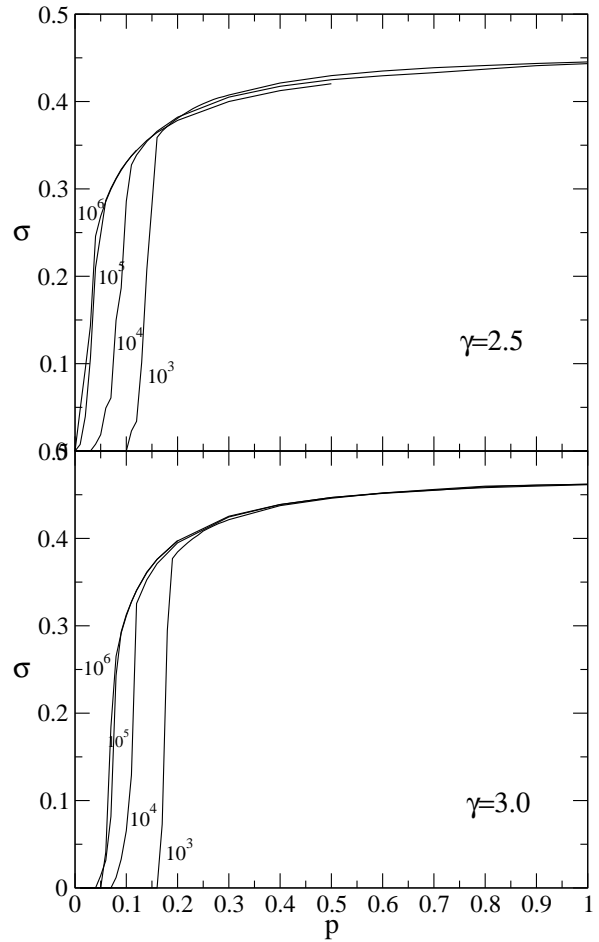}
\end{center}
\caption{Order parameter $\sigma$ as a function of the population density $p$ for scale-free
networks of varying sizes from $N=10^3$ to $N=10^6$. Results for (a) $\gamma=2.5$ and (b) $\gamma=3.0$.
\label{fig3}}
\end{figure}

Finally, we studied the effect of the network size on our presented results (Fig.~\ref{fig3}).
The curves seem to converge for networks of $N=10^5$-$10^6$ nodes.
For a given $\gamma$ value, all network sizes used follow roughly a common curve at large population densities.
The transition threshold, on the other hand, varies with $N$. Increasing the network size leads to a lower
transition value $p_c$. 
The value of $p_c$ for $\gamma=2.5$ tends to 0, for large enough networks, while for $\gamma=3.0$ it
tends to a small value of around $p_c=0.04$. Inspection of other $\gamma$ values indicates that in
the range $2<\gamma<3$ the value of $p_c$ tends to zero with increasing network size,
while when $3<\gamma<4$ the transition point is around $p_c=0.05$.

\section{Conclusions}

In this work we studied the model of Bonabeau for the case where the population moves on the nodes
of a scale-free network. A number of important differences were observed, as compared to the case of
lattice diffusion. The heterogeneity of the scale-free structure and the different behavior of the
diffusion process strongly affect the results of the model. For example, it is known that diffusion
is not a very efficient process on networks, in the sense that a random walker can never really get
away from the origin on finite-size networks \cite{Gallos,benAvraham}. This factor causes the individuals
to remain close to each other and a large number of encounters take place, even there are only few individuals.
This results in a extremely low value of the density threshold that separates egalitarian from hierarchical societies.
In fact, for $\gamma<3$ there is a strong indication from the simulations that $p_c=0$, at least for large network sizes $N$.

The number of individuals with strong probability of winning is also significantly lower than the
number of people that cannot easily win a fight and thus climb in the hierarchy. This assymetry is
not observed in lattices, where the isotropic environment of motion equally favors the development of the two
separated communities, but with equal number of members.

\section*{Acknowledgments}
The author acknowledges financial support from a European research NEST project DYSONET 012911.


\end{document}